\documentclass[twocolumn,twoside,slac]{revtex4}
\usepackage{graphicx}
\usepackage{amsmath}
\usepackage{fancyhdr}
\begin{document}

\title{Lattice QCD Production on Commodity Clusters at Fermilab}

%

\author{D. Holmgren, A. Singh, P. Mackenzie, J. Simone}
\affiliation{Fermi National Accelerator Laboratory, Batavia, IL 60510, USA}
\author{Steven Gottlieb}
\affiliation{Indiana University, Bloomington, IN 47405, USA}

\begin{abstract}
We describe the construction and results to date of Fermilab's three
Myrinet-networked lattice QCD production clusters (an 80-node dual Pentium III
cluster, a 48-node dual Xeon cluster, and a 128-node dual Xeon cluster).  We
examine a number of aspects of performance of the MILC lattice QCD code
running on these clusters.
\end{abstract}

\maketitle

\thispagestyle{fancy}


\section{Introduction}
Large scale QCD Monte Carlo calculations have typically been performed on
either commercial supercomputers or specially built massively parallel
computers.  Commodity clusters equipped with high performance networking
equipment present an attractive alternative, achieving superior performance to
price ratios and offering clear upgrade paths.  The U.S. Department of Energy,
through the SciDAC (Scientific Discovery through Advanced Computing) program,
is supporting the investigation of commodity clusters as well as purpose built
machines for lattice gauge computations.

Lattice QCD codes are memory bandwidth, floating point, and network intensive.
Successful design of clusters to run these codes requires knowledge of the
bottlenecks which control performance.  In this paper we examine a number of
aspects of performance, observed while running the MILC lattice QCD code.  For
single systems, we discuss the effects of the various memory architectures.
We examine optimizations such as data layout and SSE-assisted matrix algebra,
and consider SMP behavior.  We discuss scaling of the MILC code on
Myrinet-connected clusters.  Using a modified version of GM which allows
control of bandwidth and latency, we examine the sensitivity of the code to
network performance.

\section{The SciDAC Lattice Gauge Computing Initiative}
In 2001, a collaboration which includes most U.S. lattice theorists, applied
for funding to the DOE through the SciDAC initiative~\cite{scidac}.  Support
for fiscal years FY2002 through FY2004 was awarded, with funding primarily for
software development, but also for prototype clusters.  The SciDAC Lattice
Gauge Computing Project~\cite{scidac_lqcd} pursues hardware investigations
along two paths: special purpose computers (QCDOC~\cite{qcdoc}) under
development by Columbia University, and commodity clusters.  A primary goal of
the project is to design and implement the software infrastructure necessary
to allow legacy and new lattice gauge codes to run on both types of hardware.
At the present time it is anticipated that purpose built machines like the
QCDOC will achieve the best performance to price ratio for several years, with
clusters taking the lead in the following years.  Ideally the U.S. community
will develop and have access to several facilities which will each provide
sustained computing performance on large problems of order one to ten
teraflops.

SciDAC software~\cite{scidac_soft} will include low level application program
interfaces for communications (QMP) and for linear algebra (QLA).  QMP will
offer features similar to a small subset of MPI, with less overhead.
It will run transparently over the QCDOC mesh, Myrinet GM, gigabit ethernet
meshes, and MPI.  QLA will perform linear algebra math kernels which will run
on a single node; it will be lattice aware, offering operations at single
sites and across the entire sublattice held by a single computational node.
QLA will have low level optimizations for some architectures, for example,
through the use of the SSE (Streaming SIMD Extension) instructions on x86
CPUs.  Lattice wide computations, which involve communications, will be
addressed by the QDP API.  Lattice input and output, including parallel file
I/O when allowed by hardware, will be provided by the QIO API.

\section{The Fermilab Clusters}
Shown in Table~\ref{t_clusters} are the parameters of the lattice QCD clusters
at Fermilab.  All of these systems were purchased with supplemental grants
from the Department of Energy, the latter three under the SciDAC program.  On
all clusters we use the OpenPBS batch software with the Maui scheduler to
control user jobs.  The MPICH implementation of MPI with the Myricom-supplied
driver is used for the communications API; currently we use version 1.2.4..8a,
built against GM version 1.6.3.

\begin{table*}[htb]
\begin{center}
\begin{tabular}{|l||l||l||l||l|}
\hline
 & {\bf Pentium III} & {\bf Xeon \#1} & {\bf Xeon \#2} & {
\bf Itanium2 } \\
\hline
{\textbf{Speed}} & 700 MHz & 2.0 GHz & 2.4 GHz & 900 MHz \\
\hline
{\textbf{\# Nodes}} & 80 & 48 & 128 & 8 \\
\hline
{\textbf{\# Processors}} & 160 & 96 & 256 & 10 \\
\hline
{\textbf{Memory}} & 256 MB SDRAM & 1 GB DDR200 & 1 GB DDR200 & 1 GB DDR266 \\
\hline
{\textbf{Chipset}} & 440GX & E7500 & E7500 & zx1 \\
\hline
{\textbf{Myrinet}} & LANai-9 Copper & LANai-9 Fiber & LANai-9 Fiber & LANai-7 LAN \\
\hline
{\textbf{Other Network}} & & GigE Mesh (16) & & Dolphin SCI \\
\hline
{\textbf{Vendor}} & SGI (VA Linux) & SteelCloud & CSI & HP \\
\hline
{\textbf{Funding}} & DOE & {SciDAC} & {SciDAC} & {SciDAC} \\
\hline
{\textbf{Date in Service}} & Jan 2001 & July 2002 & Jan 2003 & May 2003 \\
\hline
\end{tabular}
\caption{\label{t_clusters}Fermilab Lattice QCD Clusters}
\end{center}
\end{table*}

As is traditional with MPICH, user jobs are launched on the set of nodes
assigned by the batch queue system by means of a user script started on the
rank 0 node.  The user script invokes {\bf mpirun}, which uses {\bf rsh} to
start the user binary on each of the nodes.  Unfortunately, there is a 256
process barrier encountered for large jobs.  Because standard {\bf rsh} uses
a pair of privileged TCP ports to connect to the target node, and privileged
ports are limited in number, large jobs fail because of exhaustion of these
TCP ports.  Instead of {\bf rsh}, use instead use {\bf ssh}, which can be
configured to use non-privileged ports.  To avoid burdening users with the
task of managing ssh keys, we also configure the ssh daemons to accept RSA
host authentication.

\section{\label{Single Node Performance}Single Node Performance}
\begin{figure}[htb]
\centering
\includegraphics[width=80mm]{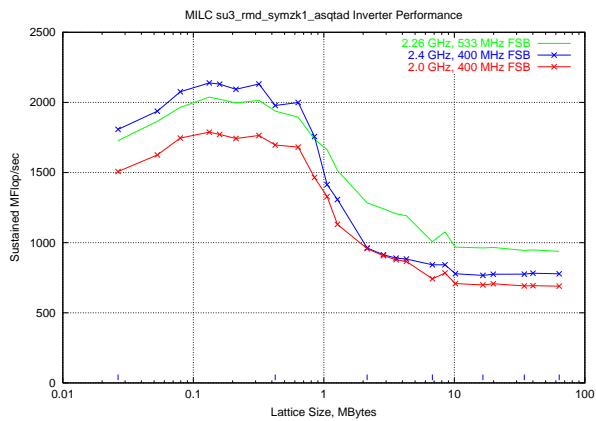}
\caption{\label{single_sample}Single Node Performance.  Each lattice site is 1656 bytes in size.
The tick marks along the abscissa mark lattices of size n$^4$, where
n = (2,4,6,8,10,12,14).}
\end{figure}

For several years we have used the performance of the Dirac inverter in the
MILC improved staggered code (``su3\_rmd\_symzk1\_asqtad'') to benchmark the
performance of systems.  Figure~\ref{single_sample} is a typical
performance graph, which shows the sustained floating point operations
delivered by different processors as a function of the lattice size, when the
MILC code is run as a single process on a single computer.  The performance
for lattices smaller than 4$^4$ is dominated by the floating point power of
the processor.  Near 4$^4$ on the processors shown, the lattice begins to
surpass the L2 cache size (512K bytes), and for larger lattices the memory
bandwidth of the computer determines performance.  The processor with the
fastest front side bus in the graph has the highest performance in main
memory, even though it does not have the highest clock speed.

To help assess floating point performance, we have implemented a modified
version of McCalpin's ``Streams'' benchmark~\cite{streams}.  The new
benchmark, qcdstream, uses several SU3 math kernels with a variety of memory
access patterns~\cite{qcdstream}.  Table~\ref{t_qcdstream} shows the
performance of two of the kernels on systems using 2.0 GHz Xeon processors
based on the E7500 chipset (interleaved DDR memory).  Also shown is the
measured performance of 900 MHz Itanium 2 processors on the same kernel.  The
Xeon processors ran standard ``C'' language code, optimized by the GCC
compiler.  The Itanium 2 processors ran kernels optimized by John Dupuis of
Hewlett-Packard.  The memory access patterns are as follows: ``in-cache'' -
operands are reused repeatedly; ``sequential'' - operands are assigned from
consecutive locations in pre-allocated memory regions; ``strided'' - operands
are assigned from non-adjacent locations from pre-allocated memory regions as
determined by a stride constant significantly larger than a cache line;
``mapped'' - operands are assigned from pre-allocated memory regions via a
randomly assigned indirect mapping.

\begin{table*}[htb]
\begin{center}
\begin{tabular}{|l|c|c|c|}
\hline
\textbf{Access Pattern} & \textbf{Matrix-Vector (MFlop/Sec)} & \textbf{Matrix-Matrix (MFlop/sec)} \\ \hline
{\bf In Cache} & 905  & 954 \\
{\bf Sequential} & 710 {(1553)}  & 815 {(2590)} \\
{\bf Strided} & 139 {(540)}  &  292 {(1326)}\\
{\bf Mapped} & 131 {(483)} & 265 {(1202)}\\ \hline
\end{tabular}
\caption{\label{t_qcdstream}Floating Point Performance in Main Memory.  Shown
are the result of the {\em qcdstream} benchmark, as measured on a 2.0 GHz Xeon
processor.  Shown in parenthesis are results for a 900 MHz Itanium 2
processor.}
\end{center}
\end{table*}

The memory bandwidth of the commodity systems we have investigated is a
function of both the processor and the chipset.  For x86-type processors, the
highest memory bandwidths currently available are on Pentium 4 systems with
800 MHz front side buses with interleaved DDR chipsets, such as Intel's i875.
Table~\ref{t_membw} lists memory bandwidths as measured by the ``copy'' section
of McCalpin's Stream benchmark.

\begin{table*}[htb]
\begin{center}
\begin{tabular}{|l|l|l|c|}
\hline
\textbf{Processor} & \textbf{Memory Type} & \textbf{Chipset} & \textbf{Bandwidth MB/sec} \\ 
\hline
{\bf Pentium} III & 100 MHz SDRAM & 440GX & 330 \\ \hline
{\bf Athlon} & DDR200 & 760MP & 700 \\ \hline
{\bf Xeon} &  DDR200 & GC-HE & 935 \\
 & DDR200 & E7500 & 1240 \\
 & DDR266 & E7501 & 1506 \\
 & PC800 RDRAM & i860 & 1305 \\ 
 & (SSE assist) & i860 & 2121 \\ \hline
{\bf Pentium 4} & PC800 RDRAM & i850 & 1320 \\
 & PC1066 RDRAM & i850E & 2035 \\ \hline
{\bf Itanium 2} & DDR266 & zx1 & 2460 \\ \hline
\end{tabular}
\caption{\label{t_membw}Measured Memory Bandwidths.  The ``SSE Assist'' data were taken using
Intel SSE instructions which bypass the L2 cache on writing; this saves a read
memory access on each write.}
\end{center}
\end{table*}

Two optimizations are very effective in improving the performance of the MILC
inverter.  First, re-ordering the data structures so that they are ``field
major'' rather than ``site major'' has the effect of packing SU3 matrices
together and achieving more efficient memory bus utilization.  This is
particularly important on Pentium 4 class processors.  These processors have
64 byte cache lines.  Single precision SU3 matrices are 72 bytes in length, so
two cache lines must be loaded for a single matrix operand.  If the extra 56
bytes loaded are not used before they are flushed out of the L2 cache,
effective memory bandwidth is greatly reduced.  Field major ordering,
which essentially places the next matrix to be used immediately after a given
matrix in memory, substantially increases performance.

The second optimization is the use of SSE instructions to vectorize the SU3
matrix algebra.  Inline versions of the fundamental MILC kernels used in the
Dirac inverter have been implemented~\cite{sse}.  Because code which utilizes
GCC assembler macros is very difficult to debug and maintain, we instead have
written these routines using a conventional assembler ({\bf
NASM}~\cite{nasm}).  The assembly code is translated to inline GCC assembler
macros by simple parser implemented in Perl.

\begin{figure}[htb]
\centering
\includegraphics[width=80mm]{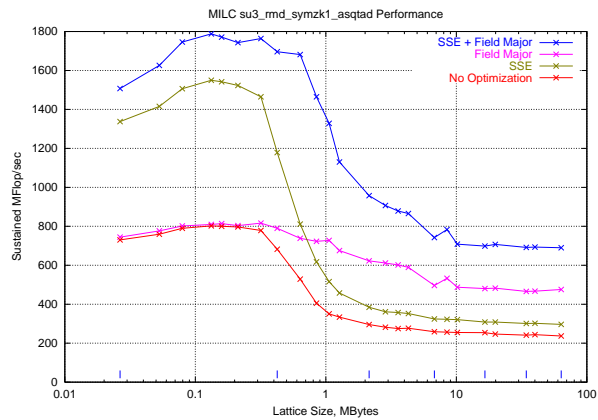}
\caption{\label{allopts}Effect of Field Major and SSE Optimizations.  All measurements were
taken on a 2.0 GHz Xeon system with the E7500 chipset.}
\end{figure}

The effects of these optimizations are shown in Figure~\ref{allopts}.  We note
that in main memory, the field major optimization greatly enhances the effect
of the SSE optimization.

\begin{figure}[htb]
\centering
\includegraphics[width=80mm]{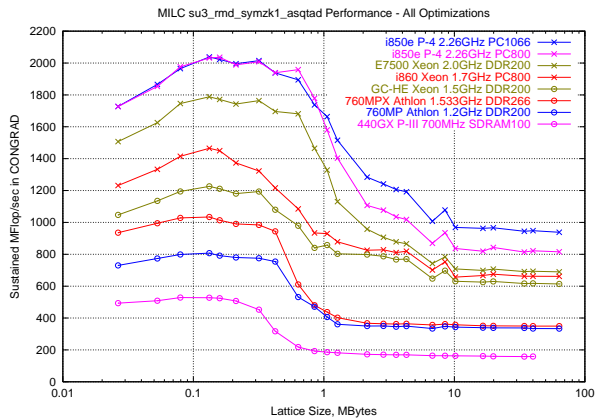}
\caption{\label{cpus}Performance Survey of Commonly Available Processors.}
\end{figure}

In Figure~\ref{cpus} are shown the measured performances of the fully
optimized code run on a variety of processors.  The top performer is the 533
MHz front side bus version of the Pentium 4.  At the time of these
measurements, Xeon systems were limited to 400 MHz front side buses.  However,
533 MHz FSB Xeons and 800 MHz FSB Pentium 4 systems are now available.  Note
that Pentium 4 systems have limited I/O buses, with only 32 bit, 33 MHz PCI
slots, whereas Xeon systems typically have one or more PCI-X buses (133 MHz,
64 bits).

\section {SMP Performance}
\begin{figure}[htb]
\centering
\includegraphics[width=80mm]{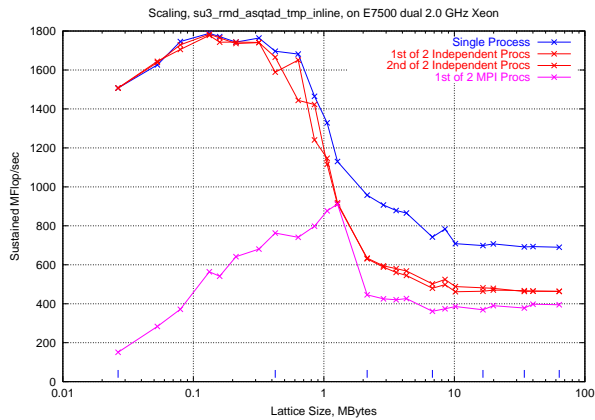}
\caption{\label{smpsingle}SMP Performance.}
\end{figure}

Commodity systems based on x86 processors are available in either single or
dual processor form.  Because high performance network adapters, such as
Myrinet, are expensive, SMP systems may be more cost effective than single
processor systems if codes scale well and the communications bandwidth
supplied by a single network interface is sufficient.  Unfortunately, as shown
in Section~\ref{Single Node Performance}, lattice codes are often memory
bandwidth bound, and dual processor x86 systems generally have the same
available bandwidth as single processor systems.  For the MILC code, this
effect is shown in Figure~\ref{smpsingle}.  The scaling of the MILC code on a
dual Xeon processor system was found to be about 65\% with two independent
processes, and about 55\% with two cooperative (MPI) processes.

Pentium 4 class processors have the best single node performance for MILC
codes.  For these processors, only SMP motherboards based on the Xeon
processor have fast, wide PCI buses, which are critical for I/O performance.
These systems can be operated with only a single processor.  However, since
the incremental cost of the second processor is a small fraction of the total
cost of the systems including the high performance network fabric, we elected
to purchase the second processor for each node of our clusters.  We note that
a number of other applications run by users on our clusters are not as
sensitive to memory bandwidth and so benefit from the availability of the
second processor.

\section{Cluster Performance}

\begin{figure}[htb]
\centering
\includegraphics[width=80mm]{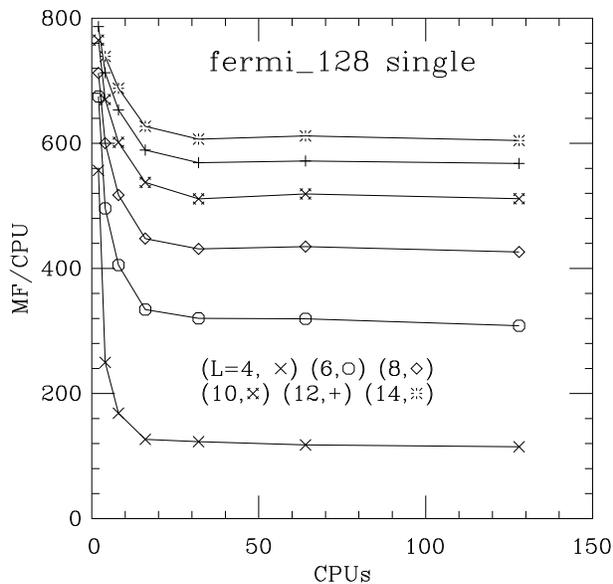}
\caption{\label{nonsmp}Cluster Performance with One Process per Node.  The distinct curves
correspond to sublattice sizes of L$^4$, where L=4,8,10,12,14.}
\end{figure}

\begin{figure}[htb]
\centering
\includegraphics[width=80mm]{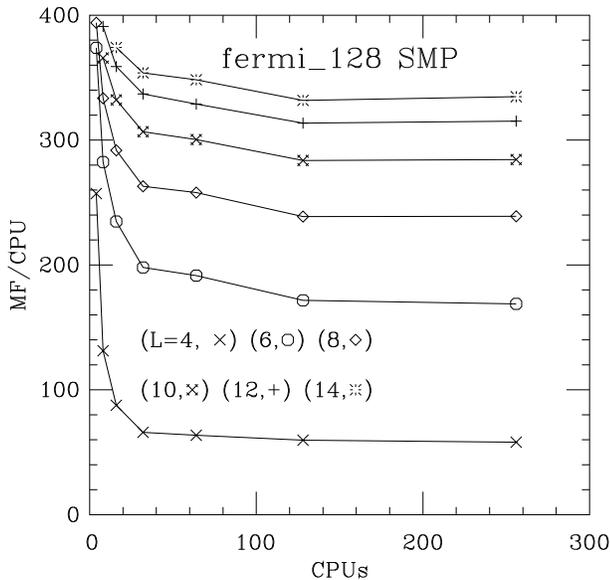}
\caption{\label{smp}Cluster Performance with Two Processes per Node.  The distinct curves
correspond to sublattice sizes of L$^4$, where L=4,8,10,12,14.}
\end{figure}

In Figures~\ref{nonsmp} and \ref{smp} we show the performance of the MILC code
on the 2.4 GHz Xeon Fermilab cluster as the number of nodes assigned to a
parallel run varies from 1 to 128.  On these runs, the sublattice assigned to
each process was kept constant.  The number of dimensions with communications
varies with the number of processes.  Communications along 1, 2, 3, and 4
directions occurs, respectively, for runs with 2, 4, 8, and 16 or greater
nodes.

The performance of the PCI bus on x86 systems varies with chipset.
Unfortunately, the best Xeon chipset for memory bandwidth, the Intel i860, has
poor I/O performance.  In Table~\ref{t_pci}, we list the burst transfer rates
between a Myrinet interface and main memory, as measured by the GM driver.
Since the Myrinet ``wire rate'' is 250 MBytes/sec, I/O performance will be
constrained on motherboards with PCI burst transfer rates below this rate.
Indeed, the large message size transfer rate asymptotes of i860 and E7500
chipset systems are, respectively, approximately 190 and 225 MBytes/sec, as
measured by the Netpipe benchmark~\cite{netpipe}.

\begin{table*}[htb]
\begin{center}
\begin{tabular}{|l|c|c|c|}
\hline
\textbf{Processor} & \textbf{Chipset} & \textbf{Bus Read (MByte/sec)} & \textbf
{Bus Write (MByte/sec)} \\ \hline
2.26 GHz Pentium 4 & i850E & 100 & 128 \\ 
700 MHz Pentium III & 440GX & 125 & 127 \\
1.7 GHz Xeon & i860 & 219 & 294 \\
2.0 GHz Xeon & E7500 & 423 & 476 \\ 
{1.7 GHz Xeon} & E7500 & 422 & 477 \\
\hline
\end{tabular}
\caption{\label{t_pci}PCI Performance of Common Motherboards}
\end{center}
\end{table*}

\begin{figure}[htb]
\centering
\includegraphics[width=80mm]{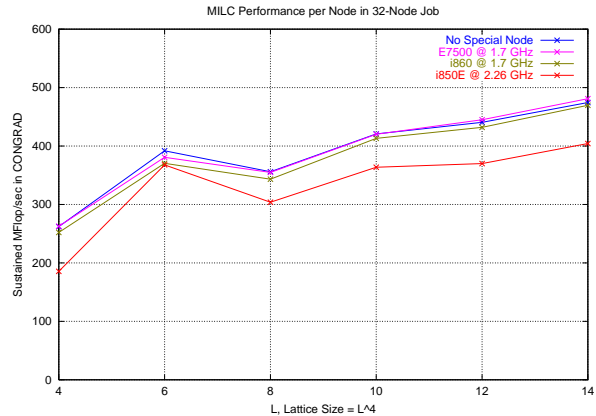}
\caption{\label{bw32}Estimating Performance of Slower Systems}
\end{figure}

To investigate the effect on lattice codes of PCI chipsets, we measured the
performance of the Dirac inverter on 32-node runs.  Since frequent barrier
synchronizations occur during the global sums used in the inverter, by
substituting just one slower system for one of the computers, we can closely
estimate the performance of a cluster consisting of slower systems.  For
this study we used the 2.0 GHz Xeon cluster based on the E7500 chipset.  We
measured performance with three slower system substitutions: a 1.7 GHz
i860-based system, a 1.7 GHz E7500-based system, and a 2.26 GHz i850E-based
system.  The latter computer has only a narrow, slow PCI bus.  The results are
shown in Figure~\ref{bw32}.  The i860-based system does not substantially
lower performance, whereas the slow, narrow PCI bus of the i850E system has a
large effect.

\begin{figure}[htb]
\centering
\includegraphics[width=80mm]{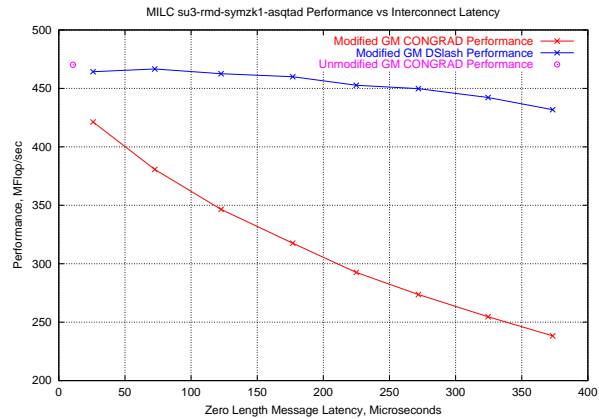}
\caption{\label{latency}Estimating Sensitivity to Communications Latency}
\end{figure}

In addition to bandwidth studies, we also investigated the effects of latency
on the MILC code.  We started with a modified version of the GM driver for
Myrinet from D. K. Panda et al~\cite{panda2}.  His group's modifications add
quality of service functionality to GM, restricting the bandwidth available to
a given connection by delaying each Myrinet packet.  We added the ability to
delay the initial packet, rather than subsequent packets, allowing control
over the latency of communications.  The results of a series of MILC runs are
shown in Figure~\ref{latency}, where we have plotted the performance of the
entire Dirac inverter (``CONGRAD'') and the underlying Dirac operator
(``D-slash'') against the zero-length message latency.  D-slash includes only
the large message communications required to access a neighboring node's
surface lattice sites.  CONGRAD includes D-slash as well as the global sums
required to test convergence.  D-slash is insensitive to latency; the full
CONGRAD, because of the global sums, is much more sensitive.

\section{Future Work at Fermilab}
Lattice gauge codes require networks with good bandwidth and low latencies.
Commercial solutions, such as Myrinet, SCI and Quadrix, are readily available
but at substantial cost.  Indeed, the fraction of the total cost of the three
Intel-based lattice gauge clusters at Fermilab represented by the Myrinet
fabrics has been about 45\%.  Gigabit ethernet offers a significantly lower
priced alternative.  Unfortunately, compared to the other networks, switched
gigabit ethernet suffers from lower bandwidth, higher latencies (because
switches typically are of the ``store and forward'' type), limited switch
sizes, and immature software (low overhead and reduced software latencies
require the use of non-TCP/IP protocol stacks such as VIA).  Gigabit ethernet
meshes, such as studied by Fodor et al~\cite{fodor}, are very inexpensive in
comparison to switched networks.  However, clusters built with these meshes
have rigid configurations, with lattice layouts fixed by wiring.  They are
also not tolerant of node failures; switched networks handle failures with
ease.  Finally, although communications with neighbor nodes have low latency
with protocols like VIA, non-nearest-neighbor communications require
interaction with the operating system of the intervening nodes, giving poor
latency performance.

Recently several FPGA (field programmable gate array) manufactures have
introduced so-called ``platform'' gate arrays, which include substantial I/O
capabilities and in some cases embedded PowerPC processors.  Fermilab has
investigated these FPGAs for applications in data acquisition, building
prototype PCI interface cards with as many as eight bidirectional high speed
serial links.  We will study the application of these cards to lattice gauge
computing.  The long term goal is to assess the feasibility of using these
interfaces to build 4-D or higher order meshes.  Like gigabit ethernet meshes,
FGPA-based meshes would provide high, scalable bandwidth (each link will have
peak bandwidth of 2 GByte/second in each direction).  Unlike gigabit ethernet
meshes, non-nearest-neighbor communications would not incur substantial
latency penalties, as routing would be performed within the fabric itself.

During FY2004, Fermilab will purchase an additional cluster.  Depending
upon final funding and the choice of CPU architecture and network fabric, this
cluster will be as large as 256 nodes.  The choices available for the
processor including Xeons with 533 MHz or higher front side buses, the new
Pentium 4 processor (``Prescott'', which includes additional SSE instructions
designed for complex arithmetic) with 800 MHz or higher front side bus, the
next generation Itanium processor (``Madison''), AMD Opteron, and the
PowerPC~970.  As discussed in this paper, dual Xeon motherboards suffer from
inadequate total memory bandwidth, but have good I/O buses.  Pentium 4
motherboards have superior memory bandwidth, but to date have only been
available with slow, narrow PCI buses.  There is hope that new Pentium 4
motherboards with high performance I/O buses (``PCI Express'') will be
available in time.  Itanium processors offer superior performance, but only
when considerable manpower is expended in hand tuning the code.  Both Opteron
and PPC970 offer the promise of superior, scalable memory bandwidth and
excellent floating point capabilities.  We will choose the processor with the
best system performance to price ratio.  For network fabrics, we will choose
between Myrinet, gigabit ethernet over Myrinet, gigabit ethernet mesh, and
Infiniband.

\begin{acknowledgments}
Work supported by Universities Research Association Inc. under Contract
No. DE-AC02-76CH03000 with the United States Department of Energy.
\end{acknowledgments}


\end{document}